\documentclass[letterpaper]{article} 
\usepackage{aaai2026}  
\usepackage{times}  
\usepackage{helvet}  
\usepackage{courier}  
\usepackage[hyphens]{url}  
\usepackage{graphicx} 
\usepackage{natbib}  
\usepackage{caption} 
\usepackage{algorithm}
\usepackage{algorithmic}
\usepackage{subfigure}
\usepackage{multirow}
\usepackage{amsmath}
\usepackage{amssymb}
\usepackage{booktabs}
\usepackage{enumitem}
\usepackage{subcaption}
\usepackage{cleveref}
\usepackage{comment}
\usepackage{xcolor}
\usepackage{colortbl}
\usepackage{nccmath}

\usepackage{verbatim}
\usepackage{soul}
\usepackage{adjustbox}
\usepackage{newfloat}
\usepackage{listings}
\DeclareCaptionStyle{ruled}{labelfont=normalfont,labelsep=colon,strut=off} 
\floatstyle{ruled}
\newfloat{listing}{tb}{lst}{}
\floatname{listing}{Listing}
\nocopyright
\title{Multimodal Representation-disentangled Information Bottleneck for Multimodal Recommendation}
\author{
    Hui Wang\textsuperscript{\rm 1},
    Jinghui Qin\textsuperscript{\rm 2}\thanks{Corresponding authors},
    Wushao Wen\textsuperscript{\rm 1}\footnotemark[1],
    Qingling Li\textsuperscript{\rm 1}, Shanshan Zhong\textsuperscript{\rm 1}, Zhongzhan Huang\textsuperscript{\rm 1}\\
}
\affiliations{
    \textsuperscript{\rm 1}Sun Yat-sen University\\
    \textsuperscript{\rm 2}Guangdong University of Technology\\

}

\usepackage{bibentry}

\begin{document}

\maketitle

\begin{abstract}
Multimodal data has significantly advanced recommendation systems by integrating diverse information sources to model user preferences and item characteristics. However, these systems often struggle with redundant and irrelevant information, which can degrade performance. Most existing methods either fuse multimodal information directly or use rigid architectural separation for disentanglement, failing to adequately filter noise and model the complex interplay between modalities. To address these challenges, we propose a novel framework, the Multimodal Representation-disentangled Information Bottleneck (MRdIB). 
Concretely, we first employ a Multimodal Information Bottleneck to compress the input representations, effectively filtering out task-irrelevant noise while preserving rich semantic information. Then, we decompose the information based on its relationship with the recommendation target into unique, redundant, and synergistic components. We achieve this decomposition with a series of constraints: a unique information learning objective to preserve modality-unique signals, a redundant information learning objective to minimize overlap, and a synergistic information learning objective to capture emergent information. By optimizing these objectives, MRdIB guides a model to learn more powerful and disentangled representations. Extensive experiments on several competitive models and three benchmark datasets demonstrate the effectiveness and versatility of our MRdIB in enhancing multimodal recommendation.
\end{abstract}


\section{Introduction}
The proliferation of multimodal data has transformed recommendation systems, particularly Multimodal Recommendation Systems (MRSs), which harness information from diverse modalities like text and images to model user preferences and item characteristics~\cite{zhou2019multi,lei2023learning}. By integrating these heterogeneous data sources, MRSs can achieve a more comprehensive understanding of user-item interactions, leading to improved recommendation accuracy~\cite{zhou2019multi,lei2023learning}.

\begin{figure}[t]
    \centering
    \includegraphics[scale=0.35]{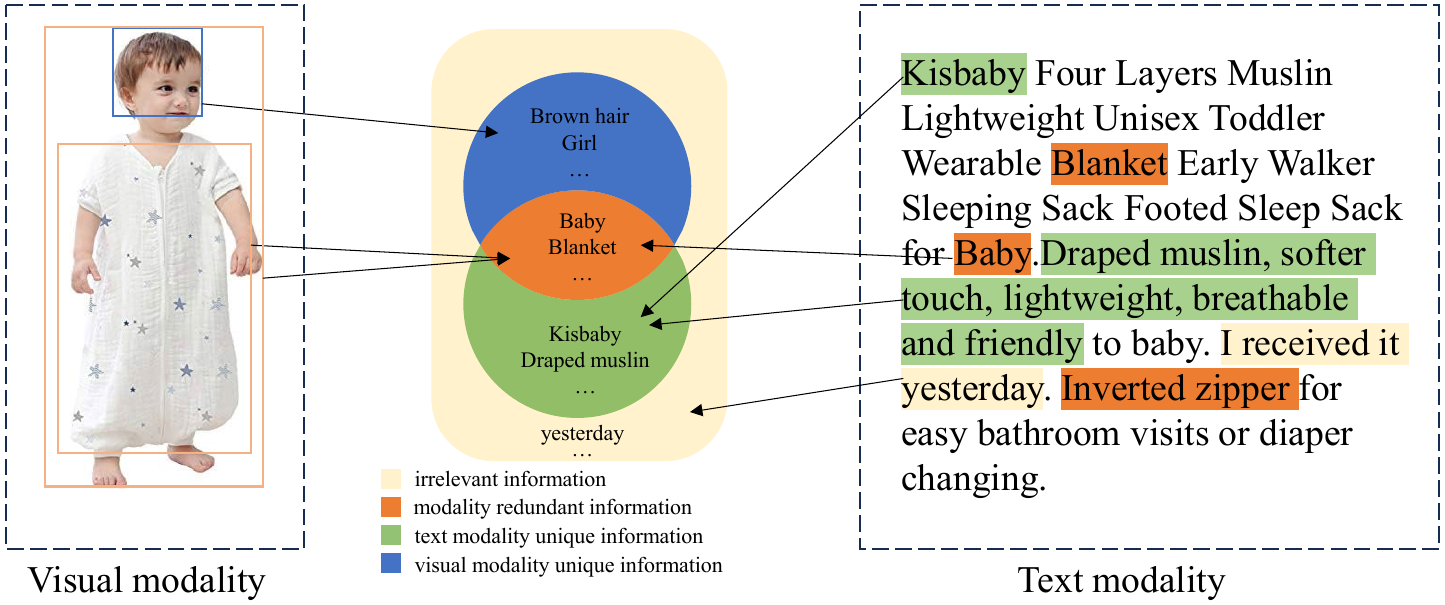}
    \vspace{-3mm}
     \caption{An example from the Amazon-baby dataset, illustrating the irrelevant information (yellow), redundant information (orange), and modality-unique information (blue and green) across different modalities.}
     \label{fig:infro_fig4}
     
\end{figure}

Despite their potential, MRSs face a fundamental challenge in learning effective representations from complex multimodal information. \textbf{First, they struggle with the effective fusion of information, as modalities are often plagued by task-irrelevant information and noise}~\cite{mo2024multimodal, liu2022disentangled,liu2023multimodal,li2025teach,li2024multimodal}. As illustrated in Figure~\ref{fig:infro_fig4}, multimodal data is often replete with noise (e.g., background clutter in an image) that is unrelated to the recommendation task. Many existing MRSs~\cite{zhou2023tale, yu2023multi, zhou2023comprehensive} attempt to fuse raw or pre-purified modality features directly. However, without an explicit mechanism to compress and discard irrelevant information, these methods can contaminate the learned representations, leading to suboptimal performance. The counter-intuitive finding in prior work~\cite{liu2022elimrec,zhang2024modality}, where single-modality models sometimes outperform their multimodal counterparts, powerfully illustrates these challenges. It suggests that without robust mechanisms to both filter noise and disentangle the complex structure of relevant information, models learn sub-optimal fused representations, where simply adding modalities can degrade, rather than improve, performance.

\textbf{Second, even after isolating task-relevant information, existing methods fail to finely disentangle its complex structure.} The relevant information is not monolithic. According to information-theoretic decomposition~\cite{kolchinsky2022novel,kolchinsky2024partial,luppi2024information}, it is a composite of \textbf{unique} components (specific to one modality), \textbf{redundant} components (shared across modalities), and \textbf{synergistic} components (emerging only from the combination of modalities). Early approaches that simply concatenate features overlook this intricate interplay entirely. More recent disentanglement methods~\cite{liu2022disentangled, zhang2024modality} attempt to address this by decomposing features into modality-unique and modality-shared representations by using separate, dedicated encoders. However, this approach has significant limitations. The rigid architectural separation is inflexible and fails to model the synergistic components that emerge only from the interaction between modalities. Crucially, these methods lack a principled, information-theoretic framework to simultaneously model and disentangle the complex and overlapping nature of multimodal information.

To overcome these limitations, we argue for a more flexible and principled approach. Instead of designing a new monolithic architecture, we propose a general plugin based on multimodal information bottleneck and multiple representation-disentangled learning objectives to extract and purify effective task-relevant information for multimodal recommendation, thus addressing the aforementioned challenges. We name this plugin Multimodal Representation-disentangled Information Bottleneck (\textbf{MRdIB}). It can be integrated into existing recommendation models to guide the model training through targeted, information-theoretic constraints. Specifically, to address the first challenge, MRdIB employs a \textbf{Multimodal Information Bottleneck} to compress the input representations, effectively filtering out task-irrelevant noise while preserving predictive information. To address the second challenge, MRdIB introduces a series of learning objectives derived from an information-theoretic decomposition of the relevant signals. The first is the \textbf{unique information learning objective}, which ensures modality-specific signals are preserved in each unimodal representation. The second one is the \textbf{redundant information disentangling objective}, which aims to minimize the informational overlap between the unimodal representations, forcing the model to learn more compact and non-overlapping features. The last one is the \textbf{synergistic information learning objective}, which encourages the fused representation to capture the extra predictive power that arises only from combining modalities. By integrating these objectives, MRdIB acts as a plug-and-play module that enables existing MRSs to learn more robust and effective multimodal representations. Extensive experiments show that MRdIB can improve the recommendation performance of a wide range of baseline models on several benchmarks, demonstrating its effectiveness and versatility.

Overall, the main contributions of this paper are 3-fold.
\begin{itemize}[leftmargin=0.4cm, itemsep=2pt, topsep=2pt] 
\item We propose a novel framework based on multimodal information bottleneck and multiple representation-disentangled learning objectives, MRdIB, to guide existing models in learning more effective multimodal representations. This provides a flexible alternative to rigid architectural changes. 
\item We ground our framework in a principled information-theoretic decomposition, addressing both the filtering of irrelevant information via an information bottleneck and the fine-grained disentanglement of unique, redundant, and synergistic information. 

\item Extensive experiments on multiple real-world datasets and baseline models demonstrate that MRdIB consistently improves recommendation performance and can be seamlessly integrated into existing MRSs, showing its versatility and effectiveness.
\end{itemize}

\section{Related Work}
\label{sec:formatting}
\subsection{Multimodal Recommendation Systems}
Early methods like VBPR~\cite{he2016vbpr} integrated visual features into matrix factorization. Subsequently, graph-based models became prominent, with works like MMGCN~\cite{wei2019mmgcn}, DualGNN~\cite{wang2021dualgnn}, and FREEDOM~\cite{zhou2023tale} exploring various graph structures to capture high-order user-item relationships. More recently, self-supervised learning has been employed to align modalities and learn robust representations, with methods like SLMRec~\cite{tao2022self} and MGCN~\cite{yu2023multi} using contrastive objectives. The latest research trends include leveraging Large Language Models (LLMs) for enhanced semantic understanding~\cite{meng2025doge}, using diffusion models for data augmentation~\cite{song2025diffcl}, and developing ID-free models like MOTOR~\cite{zhang2024learning} to address cold-start issues.

\subsection{Information Bottleneck}
The Information Bottleneck (IB) principle is a theoretical framework designed to extract the most relevant information from input data while compressing noise and irrelevant information ~\cite{tishby2000information, vera2018role, alemi2016deep}.
CGI~\cite{wei2022contrastive} uses the IB principle to filter out irrelevant information across multiple views obtained through graph augmentation during the contrastive learning process. CDRIB~\cite{cao2022cross} devises two IB regularizers to model both cross-domain and in-domain user-item interactions, resulting in unbiased representations. M3IB~\cite{du2022m} extends the IB theory from a unimodal setting to a multi-modal one, aiming to filter out irrelevant information in multimodal sequence recommendation, but it does not consider the redundant features and complementary modality-unique features, leading to redundant and inefficient representations.
\section{Problem Overview} 
\label{sec:problem}

\begin{figure*}[t]
    \centering
     \includegraphics[width=0.8\textwidth]{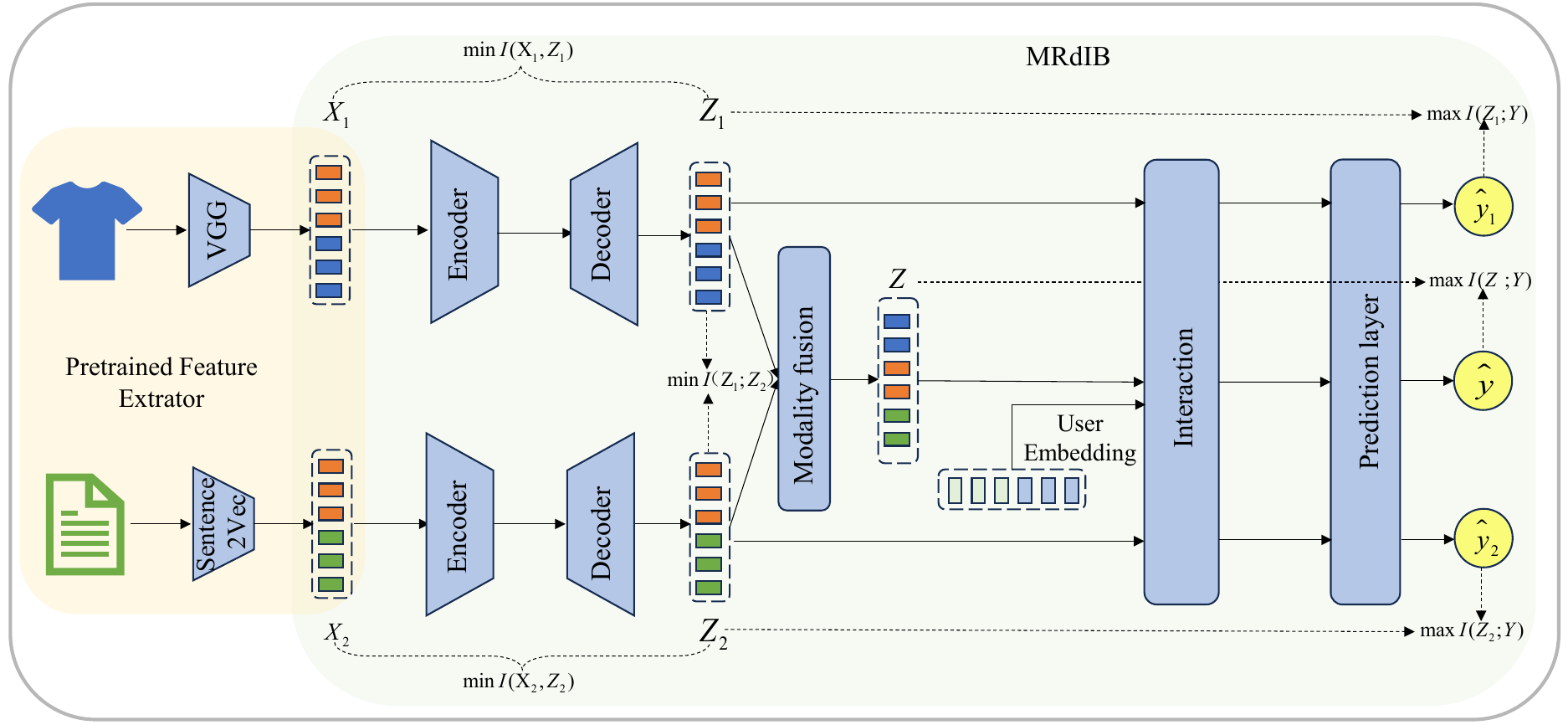}
     \vspace{-3mm}
     \caption{The overall framework of MRdIB.}
     \label{fig:MRdIB}
\end{figure*}

For the sake of simplicity, we only use two modalities in our derivations, but it can actually be easily extended to accommodate three or more modalities. We denote the input modality features as $X_1$ and $X_2$, representing the visual and textual modalities, respectively. The corresponding unimodal representations are denoted as $Z_1$ for the visual part and $Z_2$ for the textual part. $Z$ represents the fused multimodal representation that integrates information from both modalities, while $Y$ denotes the target output, typically the recommendation score. This formulation provides a structured framework for understanding the information flow in MRSs, which can be summarized as follows:
\begin{small}
\begin{equation}
    \begin{aligned}
    &  X_1 \rightarrow Z_1 \\
    &  X_2 \rightarrow Z_2
    \end{aligned} \rightarrow Z \rightarrow Y
    \label{eq:workflow}
\end{equation}
\end{small}

\noindent The quality of the learned representation $Z$ is critical. An ideal representation should be a \textit{sufficient statistic} for the target $Y$, meaning it captures all necessary predictive information, while also being \textit{minimal}, meaning it discards all task-irrelevant noise from the inputs $X_1$ and $X_2$. However, obtaining an ideal representation is non-trivial. As highlighted in the introduction section, existing methods often struggle to compress noise and are unable to properly disentangle the complex interplay of unique, redundant, and synergistic information within the task-relevant signals. This leads to suboptimal fused representations, ultimately hindering recommendation performance.

\section{Methodology}
To address the challenges of noise and information entanglement outlined above, we introduce the Multimodal Representation-disentangled Information Bottleneck (MRdIB), a framework designed to learn robust and disentangled representations. MRdIB systematically tackles both issues, as shown in Figure~\ref{fig:MRdIB}. First, to handle irrelevant information, we employ a Multimodal Information Bottleneck (MIB) that compresses the inputs and filters out noise. Second, to disentangle the structure of the relevant information, we use Partial Information Decomposition (PID)~\cite{kolchinsky2022novel,kolchinsky2024partial} to analyze its unique, redundant, and synergistic components. Finally, we translate this decomposition into a set of concrete learning objectives that guide the parameter optimization of any underlying recommendation model. 

\subsection{Filtering Irrelevant Information with MIB}
The first step in our framework is to address the challenge of noisy, task-irrelevant information. To achieve this, we employ the Multimodal Information Bottleneck (MIB) principle. The goal of MIB is to learn compressed representations, $Z_1$ and $Z_2$, from the raw inputs $X_1$ and $X_2$ that are maximally informative about the target $Y$ while being minimally informative about the inputs themselves. This forces the model to discard irrelevant details (noise) and retain only the predictive signals. This objective can be formally expressed as:
\begin{small}
\begin{equation}
\begin{split}
    \mathcal{L}_{\text{MIB}} = \min_{Z_1, Z_2}  I(X_1; Z_1) + I(X_2; Z_2) - \beta I(Z_1, Z_2; Y)
\end{split}
\label{eq:mib}
\end{equation}
\end{small}

\noindent where $I(\cdot;\cdot)$ denotes mutual information, and $\beta$ is a hyperparameter that balances the trade-off between compression and prediction. The first term, $I(Z_1, Z_2; Y)$, encourages the learned representations to be predictive of the target. The second term, consisting of $I(X_1; Z_1)$ and $I(X_2; Z_2)$, encourages the representations to be concise by minimizing the information they retain from the original inputs.

However, the mutual information terms in \Cref{eq:mib} are intractable to compute directly. We therefore derive a tractable variational upper bound for this objective, a standard technique in IB-based methods~\cite{alemi2016deep}. This involves introducing a variational encoder $q_\phi(z_i|x_i)$ to approximate the true posterior $p(z_i)$ and a generative decoder $p_\theta(y|z_1, z_2)$ to model the target likelihood. By assuming a simple prior (e.g., a standard Gaussian $\mathcal{N}(0, I)$), we can derive the following optimizable loss function (see Appendix for the full derivation):
\begin{small}
\begin{equation}
\begin{aligned}
    \mathcal{L}_{\text{MIB}} &= \mathbb{E}_{p_\theta(x_1,x_2,y)}[-\log p_\theta(y | z_1, z_2)] \\&+ \alpha_1 \sum_{i=1,2} KL(q_\phi(z_i|x_i) || p(z_i))
    \label{eq:mib_loss}
\end{aligned}
\end{equation}
\end{small}

The first term is a reconstruction loss, ensuring the representations are predictive. The second term, the compression constraint, is a KL-divergence term that regularizes the representations by pushing them towards the prior, thus enforcing compression.
\subsection{Decomposing Relevant Information via PID}
While the MIB objective effectively filters out task-irrelevant noise, it does not govern the internal composition of the remaining, task-relevant information. Without explicit guidance, the learning process for the filtered representations, $Z_1$ and $Z_2$, can be suboptimal. The model might learn redundant representations where both modalities encode the same shared information, leading to inefficiency. Conversely, it might discard valuable unique information from one modality or fail to form the synergistic information that only emerges from combining both. This lack of informational disentanglement means that simply fusing $Z_1$ and $Z_2$ remains a key challenge.
\begin{figure}
    \centering
    \includegraphics[width=0.7\linewidth]{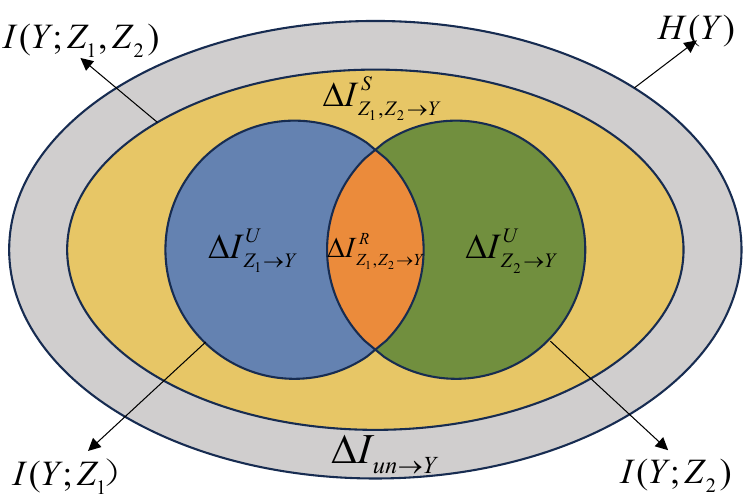}
    \vspace{-3mm}
    \caption{Multimodal Information Decomposition.}
    \label{fig:3figue}
\end{figure}
To properly disentangle these components, we turn to Partial Information Decomposition (PID)~\cite{wibral2017partial}, a theoretical framework for analyzing how a target variable $Y$ gains information from multiple sources—in our case, the multimodal representations $Z_1$ and $Z_2$. As illustrated in Figure~\ref{fig:3figue}, PID decomposes the total mutual information $I(Z_1, Z_2; Y)$ into three distinct, non-negative components:
\textbf{1) Unique Information} ($\Delta I^U_{Z_i \to Y}$): Information about $Y$ that can only be obtained from one source. For example, the aesthetic appeal of an item might be uniquely present in the image representation ($Z_1$).
\textbf{2) Redundant Information} ($\Delta I^R_{Z_1,Z_2 \to Y}$): Information about $Y$ that is shared and available from either source. For instance, the item's category might be inferred from both its image ($Z_1$) and its textual description ($Z_2$).
\textbf{3) Synergistic Information} ($\Delta I^S_{Z_1, Z_2 \to Y}$): New information about $Y$ that emerges only when both sources are considered together. For example, sarcasm in a product review might only be detectable by combining the text ($Z_2$) with a user's profile picture ($Z_1$), revealing a preference pattern not visible in either modality alone.
This decomposition provides a precise, information-theoretic blueprint for disentanglement. Instead of relying on ad-hoc architectural choices, we can use these components to define clear learning objectives. The next section details how we translate this PID framework into a concrete set of loss functions that guide the model to learn disentangled representations.
\subsection{Learning Objectives for Disentanglement}
Building on the PID framework, we introduce a set of learning objectives designed to enforce the decomposition of information within the learned representations. These objectives collectively form a representation disentanglement loss, denoted as $\mathcal{L}_{\text{Rd}}$. The goal is to guide the model to learn representations ($Z_1, Z_2$) that are not only predictive of the target $Y$, but whose informational composition is explicitly disentangled into unique, redundant, and synergistic components. The overall disentanglement loss is formulated as a weighted sum of these objectives:
\begin{small}
\begin{equation}
    \mathcal{L}_{\text{Rd}} = \mathcal{L}_{\Delta I^U_{Z_1 \to Y}, \Delta I^U_{Z_2 \to Y}} + \mathcal{L}_{\Delta I^R_{Z_1,Z_2 \to Y}} + \mathcal{L}_{\Delta I^S_{Z_1, Z_2 \to Y}}
    \label{eq:Rdobject}
\end{equation}
\end{small}

\noindent where $\mathcal{L}_{U}$ encourages the preservation of unique information, and $\mathcal{L}_{R}$ promotes the minimization of redundant information. The synergistic component is implicitly optimized through the main prediction task, as we will explain.

\subsubsection{Preserving Unique Information}
To ensure that information exclusive to each modality is not lost during compression, each unimodal representation ($Z_1$ and $Z_2$) must be independently predictive of the target $Y$. This objective forces the model to preserve modality-specific signals. We formulate this by maximizing the mutual information between each unimodal representation and the target:
\begin{small}
\begin{equation}
    \max I(Z_1; Y) + I(Z_2; Y)
\end{equation}
\end{small}

\noindent This is equivalent to minimizing the following loss term, $\mathcal{L}_{U}$, which can be tractably estimated using the negative log-likelihood of the target given each representation:
\begin{small}
\begin{equation}
    \mathcal{L}_{\scriptscriptstyle\Delta I^U_{Z_1 \to Y}, \Delta I^U_{Z_2 \to Y}} = \mathbb{E}_{p_\theta(x_1,x_2,y)}[\sum_{i=1,2}-p_{\theta}(z_i|x_i)\log p_\theta (y|z_i))]
    \label{eq:unique}
\end{equation}
\end{small}

\noindent where separate decoders $p_\theta(y|z_i)$ are used to predict the target from each modality's representation.
\subsubsection{Minimizing Redundant Information}
To address redundant information, we aim to prevent the model from encoding the same information in both representations. We achieve this by imposing a constraint that minimizes the informational overlap between $Z_1$ and $Z_2$. This encourages the representations to be compact and specialized. The objective is to minimize their mutual information:
\begin{small}
\begin{equation}
    \min I(Z_1; Z_2)
\end{equation}
\end{small}

\noindent Since directly optimizing mutual information is intractable, we design a neural network $f$ to estimate the bound of $I(Z_1; Z_2)$, following the principles of MINE~\cite{ishmael2018mine}. This approach yields the following loss term, with the full derivation available in the Appendix:
\begin{small}
\begin{equation}
    \mathcal{L}_{\Delta I^R_{Z_1,Z_2 \to Y}} = \mathbb{E}_{p(z_1, z_2)}[f(z_1, z_2)] - \log \mathbb{E}_{p(z_1)p(z_2)}[e^{f(z_1, z_2)}]
    \label{eq:redundant}
\end{equation}
\end{small}

\noindent By minimizing this loss, we force the representations towards statistical independence, thereby disentangling the shared, redundant information.

\subsubsection{Capturing Synergistic Information}
Synergistic information is the additional predictive power that emerges only when both modalities are combined. This is the most critical component for effective multimodal fusion. To ensure the model captures this, we formulate the objective as maximizing the mutual information between the joint representation and the target:
\begin{small}
\begin{equation}
    \max I(Z_1, Z_2; Y)
    \label{eq:syn}
\end{equation}
\end{small}

\noindent This constraint is mathematically equivalent to optimizing: 
\begin{small}
\begin{equation}
\label{eq:synergistic}
\mathcal{L}_{\Delta I^S_{Z_1, Z_2 \to Y}} \!=\! \mathbb{E}_{p_\theta(x_1,x_2,y)}
\big[ -p_\theta(z_1,z_2|x_1,x_2)\log p_\theta(y|z_1,z_2) \big]
\end{equation}
\end{small}

By combining the learning objectives in~\Cref{eq:unique,eq:redundant,eq:synergistic} with the MIB loss function in~\Cref{eq:mib_loss}, the final optimization objective is formulated as:
\begin{small}
\begin{equation}
\begin{aligned}
     \mathcal{L}&_{\text{MRdIB}} =\mathcal{L}_{\text{MIB}} + \mathcal{L}_{\text{Rd}}\\
     &=\mathbb{E}_{p_\theta(x_1,x_2,y)}\big[ -p_\theta(z_1,z_2|x_1,x_2)\log p_\theta(y\mid z_1,z_2) \\
    &  +\alpha_1\sum\nolimits_{i=1,2} KL(q_\phi(z_i|x_i) || p(z_i)) \\
    &  +\alpha_2(E_{p_\theta(z_1,z_2)}[f]-\log E_{p_\theta(z_1)p_\theta(z_2)}[e^f]) \\
    & +\alpha_3(\sum\nolimits_{i=1,2}-p_{\theta}(z_i|x_i)\log p_\theta (y|z_i))\big]
\end{aligned}
\label{eq:object}
\end{equation}
\end{small}

\noindent where $\alpha_1$, $\alpha_2$, and $\alpha_3$ are hyperparameters that balance the contributions of each component in the overall loss function. 

\begin{table*}[!ht]
    \centering
    \small
    \resizebox{1\textwidth}{!}{
    \begin{tabular}{lcccccccccccc}
        \toprule
 & \multicolumn{4}{c}{\textbf{Baby}} & \multicolumn{4}{c}{\textbf{Sports}} & \multicolumn{4}{c}{\textbf{Clothing}}  \\
       Models & \textbf{REC} & \textbf{PREC}& \textbf{MAP} & \textbf{NDCG} & \textbf{REC} & \textbf{PREC}& \textbf{MAP} & \textbf{NDCG} & \textbf{REC} & \textbf{PREC}& \textbf{MAP} & \textbf{NDCG} \\
        \midrule
        VBPR & 0.0243 & 0.0054 & 0.0132 & 0.0163 & 0.0344 & 0.0077 &0.0184 & 0.0229 &  0.0191 & 0.0040 &0.0106 & 0.0128 \\
        VBPR+MRdIB & \textbf{0.0299} & \textbf{0.0067} & \textbf{0.0156} & \textbf{0.0196} & \textbf{0.0374} & \textbf{0.0082} & \textbf{0.0203} & \textbf{0.0250} & \textbf{0.0243} & \textbf{0.0051} & \textbf{0.0128} & \textbf{0.0158}\\
        Improv.& {+15.23\%} & {+16.67\%} & {+16.67\%} & {+16.56\%} & {+8.72\%} & {+6.49\%} & {+10.33\%} & {+9.17\%} & {+27.23\%} & {+27.50\%} & {+20.75\%} & {+23.44\%}\\
        \midrule
        MMGCN & 0.0243 & 0.0055 & 0.0126 & 0.0159 & 0.0235 & 0.0053 & 0.0117 & 0.0150 & 0.0135 & 0.0028 & 0.0068 & 0.0085 \\
        MMGCN+MRdIB & \textbf{0.0261} & \textbf{0.0058} & \textbf{0.0136} & \textbf{0.0171} & \textbf{0.0273} & \textbf{0.0061} & \textbf{0.0140} & \textbf{0.0177} & \textbf{0.0143} & \textbf{0.0030} & \textbf{0.0074} & \textbf{0.0092} \\
        Improv. & {+7.4\%} & {+5.45\%} & {7.94\%} & {+7.55\%} & {+16.17\%} & {+15.09\%} & {+19.66\%} & {+18.00\%} & {+5.93\%} & {+7.14\%} & {+8.82\%} & {+8.24\%} \\
        \midrule
        DualGNN & 0.0323	&0.0072&	0.0173&0.0215	&0.0381&	0.0085&	0.0210&	0.0258	&0.0290	&0.0060&	0.0154 &	0.0189
        \\ 
        DualGNN+MRdIB &\textbf{0.0360}&\textbf{0.0080}&\textbf{0.0199}&\textbf{0.0244}&\textbf{0.0431}&	\textbf{0.0095}&	\textbf{0.0229}	&\textbf{0.0285}	&\textbf{0.0310}&	\textbf{0.0065}&	\textbf{0.0167}	&\textbf{0.0204}
 \\
        Improv. &+11.46\%&+11.11\%&+15.03\%&+13.49\%&+13.12\%&	+11.76\%&	+9.05\%&	+10.47\%	&+6.90\%	&+8.33\%	&+8.44\%	&+7.94\% \\
        \midrule
        FREEDOM & 0.0390 & 0.0086 & 0.0202 & 0.0254 & 0.0458 & 0.0101 & 0.0244 & 0.0302 & 0.0405 & 0.0084 & 0.0220 & 0.0268 \\
        FREEDOM+MRdIB &  \textbf{0.0407} & \textbf{0.0090} & \textbf{0.0212} & \textbf{0.0266} & \textbf{0.0481} & \textbf{0.0105} & \textbf{0.0257} & \textbf{0.0318} & \textbf{0.0420} & \textbf{0.0087} & \textbf{0.0227} & \textbf{0.0277}\\
        Improv. & {+4.36\%} & {+4.65\%} & {+4.95\%} & {+4.72\%} & {+5.02\%} & {+3.96\%} & {+5.33\%} & {+5.30\%} & {+3.70\%} & {+3.57\%} & {+3.18\%} & {+3.36\%}\\
        \midrule
        MGCN & 0.0392 & 0.0086 & 0.0218 & 0.0266 & 0.0474 & 0.0104 & 0.0254 & 0.0314 & 0.0428 & 0.0089 & 0.0233 & 0.0283 \\
        MGCN+MRdIB & \textbf{0.0418} & \textbf{0.0091} & \textbf{0.0228} & \textbf{0.0280} & \textbf{0.0488} & \textbf{0.0107} & \textbf{0.0260} & \textbf{0.0322} & \textbf{0.0444} & \textbf{0.0092} & \textbf{0.0240} & \textbf{0.0291} \\
        Improv. & {+6.22\%} & {+5.49\%} & {+4.39\%} & {+5.00\%} & {+2.87\%} & {+2.80\%} & {+2.31\%} & {+2.48\%} & {+3.60\%} & {+3.26\%} & {+2.92\%} & {+2.75\%} \\
        \midrule
        SOIL & 0.0415 & 0.0092 & 0.0227 & 0.0279 & 0.0509 & 0.0112 & 0.0280 & 0.0344 & 0.0452 & 0.0093 & 0.0246 & 0.0299 \\
        SOIL+MRdIB & \textbf{0.0438} & \textbf{0.0097} & \textbf{0.0232} & \textbf{0.0288} & \textbf{0.0536} & \textbf{0.0117} & \textbf{0.0289} & \textbf{0.0356} & \textbf{0.0469} & \textbf{0.0097} & \textbf{0.0255} & \textbf{0.0310} \\
        Improv. & {+5.54\%} & {+5.43\%} & {+2.20\%} & {+3.23\%} & {+5.30\%} & {+4.46\%} & {+3.21\%} & {+3.49\%} & {+3.76\%} & {+4.30\%} & {+3.66\%} & {+3.68\%} \\
        \midrule
        Avg Improv. & \textbf{+8.37\%} & \textbf{+8.13\%} & \textbf{+8.53\%} & \textbf{+8.43\%} & \textbf{+8.53\%} & \textbf{+7.43\%} & \textbf{+8.32\%} & \textbf{+8.15\%} & \textbf{+8.52\%} & \textbf{+9.02\%} & \textbf{+7.96\%} & \textbf{+8.24\%} \\
        \bottomrule
    \end{tabular}
    }
    \vspace{-3mm}
    \caption{Performance comparison of different models enhanced by MRdIB across multiple datasets.}
    \label{tab:performance_comparison}
\end{table*}

\begin{table*}[t]
    \centering
    \small
    \belowrulesep=0pt
    \aboverulesep=0pt
    \resizebox{1\textwidth}{!}{
    \begin{tabular}{@{}l|cccc|cccc|cccc@{}}
        \toprule
 & \multicolumn{4}{c|}{\textbf{Baby}} & \multicolumn{4}{c|}{\textbf{Sports}} & \multicolumn{4}{c}{\textbf{Clothing}}  \\

       Models & \textbf{REC} & \textbf{PREC}& \textbf{MAP} & \textbf{NDCG} & \textbf{REC} & \textbf{PREC}& \textbf{MAP} & \textbf{NDCG} & \textbf{REC} & \textbf{PREC}& \textbf{MAP} & \textbf{NDCG} \\
        \midrule
        VBPR & 0.0243 & 0.0054 & 0.0132 & 0.0163 & 0.0344 & 0.0077 &0.0184 & 0.0229 &  0.0191 & 0.0040 &0.0106 & 0.0128 \\
        VBPR+MRdIB &  \textbf{0.0299} & \textbf{0.0067} & \textbf{0.0156} & \textbf{0.0196} & \textbf{0.0374} & \textbf{0.0082} & \textbf{0.0203} & \textbf{0.0250} & \textbf{0.0243} & \textbf{0.0051} & \textbf{0.0128} & \textbf{0.0158}\\
        VBPR+MRdIB($\alpha_1^-,\alpha_2,\alpha_3$) &0.0280 & 0.0063 & 0.0154 & 0.0190 & 0.0374 & 0.0082 & 0.0203 & 0.0250 & 0.0243 & 0.0051 & 0.0128 & 0.0158
  \\
        VBPR+MRdIB($\alpha_1,\alpha_2^-,\alpha_3$)&  0.0298 & 0.0067 & 0.0152 & 0.0192 & 0.0362 & 0.0080 & 0.0190 & 0.0237 & 0.0241 & 0.005 & 0.0126 & 0.0156
\\
        VBPR+MRdIB($\alpha_1,\alpha_2,\alpha_3^-$)& 0.0257 & 0.0058 & 0.0141& 0.0173 & 0.0381 & 0.0083 & 0.0203 & 0.0251&0.0206&0.0043&0.0116&0.0139
\\
        VBPR+MRdIB($\alpha_1,\alpha_2^-,\alpha_3^-$)&  0.0253 & 0.0057 & 0.0134 & 0.0168 & 0.0345 & 0.0078 & 0.0188 & 0.0232&0.0201&0.0427&0.0115&0.0138
\\
        \midrule
        MMGCN & 0.0243 & 0.0055 & 0.0126 & 0.0159 & 0.0235 & 0.0053 & 0.0117 & 0.0150 & 0.0135 & 0.0028 & 0.0068 & 0.0085 \\
        MMGCN+MRdIB & \textbf{0.0261} & \textbf{0.0058} & \textbf{0.0136} & \textbf{0.0171} & \textbf{0.0273} & \textbf{0.0061} & \textbf{0.0140} & \textbf{0.0177} & \textbf{0.0143} & \textbf{0.0030} & \textbf{0.0074} & \textbf{0.0092} \\
        MMGCN+MRdIB($\alpha_1^-,\alpha_2,\alpha_3$) &0.0243&0.0054&0.0130&0.0161&0.0257&0.0058&0.0137&0.0171&0.0136&0.0028&0.0069&0.0086   \\   
        MMGCN+MRdIB($\alpha_1,\alpha_2^-,\alpha_3$)&  0.0248&0.0055&0.0129&0.0162&0.0259&0.0058&0.0138&0.0172&0.0136&0.0028&0.0072&0.0089\\
        MMGCN+MRdIB($\alpha_1,\alpha_2,\alpha_3^-$)& 0.0220&0.0050&0.0116&0.0146&0.0250&0.0053&0.0130&0.0162&0.0137&0.0029&0.0071&0.0088   \\
        MMGCN+MRdIB($\alpha_1,\alpha_2^-,\alpha_3^-$)& 0.0245&0.0055&0.0131&0.0163&0.0239&0.0054&0.0127&0.0159&0.0139&0.0029&0.0073&0.0090\\
        \midrule
        DualGNN & 0.0323	&0.0072&	0.0173&0.0215	&0.0381&	0.0085&	0.0210&	0.0258	&0.0290	&0.0060&	0.0154 &	0.0189 \\
        DualGNN+MRdIB &\textbf{0.0360}&\textbf{0.0080}&\textbf{0.0199}&\textbf{0.0244}&\textbf{0.0431}&\textbf{0.0095}&\textbf{0.0229}&\textbf{0.0285}&\textbf{0.0310}&\textbf{0.0065}&\textbf{0.0167}&\textbf{0.0204} \\
        DualGNN+MRdIB($\alpha_1^-,\alpha_2,\alpha_3$) &0.0334&0.0074&0.0181&0.0224&0.0426&0.0093&0.0231&0.0285&0.0302&0.0063&0.0161&0.0197 \\
        DualGNN+MRdIB($\alpha_1,\alpha_2^-,\alpha_3$)& 0.0338&0.0071&0.0185&0.0224&0.0430&0.0095&0.0232&0.0286&0.0293&0.0060&0.0153&0.0188\\
        DualGNN+MRdIB($\alpha_1,\alpha_2,\alpha_3^-$)& 0.0343&0.0077&0.0184&0.0229&0.0424&0.0093&0.0232&0.0285&0.0296&0.0062&0.0161&0.0196\\
        DualGNN+MRdIB($\alpha_1,\alpha_2^-,\alpha_3^-$)&0.0341&0.0072&0.0182&0.0217&0.0427&0.0090&0.0227&0.0279&0.0295&0.0062&0.0156&0.0192 \\  
        \midrule
        FREEDOM & 0.0390 & 0.0086 & 0.0202 & 0.0254 & 0.0458 & 0.0101 & 0.0244 & 0.0302 & 0.0405 & 0.0084 & 0.0220 & 0.0268 \\
        FREEDOM+MRdIB & \textbf{0.0407} & \textbf{0.0090} & \textbf{0.0212} & \textbf{0.0266} & \textbf{0.0481} & \textbf{0.0105} & \textbf{0.0257} & \textbf{0.0318} & \textbf{0.0420} & \textbf{0.0087} & \textbf{0.0227} & \textbf{0.0277} \\
        FREEDOM+MRdIB($\alpha_1^-,\alpha_2,\alpha_3$) &  0.0391 & 0.0086 & 0.0209 & 0.0259 & 0.0464 & 0.0102 & 0.0248 & 0.0307 & 0.0407 & 0.0084 & 0.0220 & 0.0268\\
        FREEDOM+MRdIB($\alpha_1,\alpha_2^-,\alpha_3$)& 0.0403 & 0.0089 & 0.0203 & 0.0257 & 0.0470 & 0.0103 & 0.0252 & 0.0311 & 0.0409 & 0.0085 & 0.0220 & 0.0268\\
        FREEDOM+MRdIB($\alpha_1,\alpha_2,\alpha_3^-$)& 0.0386 & 0.0085 & 0.0200 & 0.0251 & 0.0474 & 0.0104 & 0.0251 & 0.0312 & 0.0384 & 0.0080 & 0.0211 & 0.0256\\ 
        FREEDOM+MRdIB($\alpha_1,\alpha_2^-,\alpha_3^-$)&  0.0380&	0.0084&0.0198&0.0248&0.0474 & 0.0104 & 0.0251 & 0.0312 & 0.0384 & 0.0080 & 0.0211 & 0.0256\\
        \midrule
        MGCN & 0.0392 & 0.0086 & 0.0218 & 0.0266 & 0.0474 & 0.0104 & 0.0254 & 0.0314 & 0.0428 & 0.0089 & 0.0233 & 0.0283 \\
        MGCN+MRdIB & \textbf{0.0418} & \textbf{0.0091} & \textbf{0.0228} & \textbf{0.0280} & \textbf{0.0485} & \textbf{0.0106} & \textbf{0.0259} & \textbf{0.0321} & \textbf{0.0437} & \textbf{0.0091} & \textbf{0.0240} & \textbf{0.0291} \\
        MGCN+MRdIB($\alpha_1^-,\alpha_2,\alpha_3$) & 0.0401 & 0.0088 & 0.0215 & 0.0266 & 0.0474 & 0.0104 & 0.0256 & 0.0316 & 0.0426 & 0.0088 & 0.0234 & 0.0284  \\
        MGCN+MRdIB($\alpha_1,\alpha_2^-,\alpha_3$)& 0.0404 & 0.0088 & 0.0222 & 0.0272 & 0.0471 & 0.0104 & 0.0257 & 0.0316 & 0.0426 & 0.0088 & 0.0233 & 0.0283 \\
        MGCN+MRdIB($\alpha_1,\alpha_2,\alpha_3^-$)& 0.0417 & 0.0091 & 0.0226 & 0.0278 & 0.0480 & 0.0105 & 0.0258 & 0.0319 & \textbf{0.0435} & \textbf{0.0090} & \textbf{0.0240} & \textbf{0.0290} \\
        MGCN+MRdIB($\alpha_1,\alpha_2^-,\alpha_3^-$)&0.0401 & 0.0088 & 0.0222 & 0.0271 &0.0479&	0.0105	&0.0259&	0.0319
& 0.0434 & 0.0087 & 0.0237 & 0.0287 \\
        \midrule
        SOIL & 0.0415 & 0.0092 & 0.0227 & 0.0279 & 0.0509 & 0.0112 & 0.0280 & 0.0344 & 0.0452 & 0.0093 & 0.0246 & 0.0299 \\
        SOIL+MRdIB & \textbf{0.0438} & \textbf{0.0097} & \textbf{0.0232} & \textbf{0.0288} & \textbf{0.0536} & \textbf{0.0117} & \textbf{0.0289} & \textbf{0.0356} & \textbf{0.0469} & \textbf{0.0097} & \textbf{0.0255} & \textbf{0.0310} \\
        SOIL+MRdIB($\alpha_1^-,\alpha_2,\alpha_3$) &0.0432&0.0094&0.0227&0.0283&0.0525&0.0114&0.0290&0.0354&0.0461&0.0096&0.0252&0.0305  \\
        SOIL+MRdIB($\alpha_1,\alpha_2^-,\alpha_3$)& 0.0418&0.0092&0.0221&0.0275&0.0506&0.0111&0.0275&0.0338&0.0466&0.0097&0.0254&0.0308 \\
        SOIL+MRdIB($\alpha_1,\alpha_2,\alpha_3^-$)& 0.0427&0.0093&0.0224&0.0282&0.0532&0.0116&0.0287&0.0354&0.0467&0.0097&0.0253&0.0308 \\
        SOIL+MRdIB($\alpha_1,\alpha_2^-,\alpha_3^-$)& 0.0420&0.0092&0.0222&0.0276&0.0506&0.0111&0.0275&0.0339&0.0465&0.0096&0.0253&0.0380 \\
        \bottomrule
    \end{tabular}
     }
     \vspace{-3mm}
    \caption{Ablation study of MRdIB. The "-" means that this parameter is set to 0 effectively discarding this component.}
    \label{tab:ablation}
\end{table*}

\section{Experiments}

\subsection{Experimental Setting}
\noindent \textbf{Datasets.} In our experiments, we utilize three categories from the Amazon review dataset~\cite{mcauley2015image}, which are referred to as Baby, Sports, and Clothing throughout the paper.
Following previous work~\cite{liu2023semantic, guo2024lgmrec, yang2023modal}, we apply a 5-core setting to filter both users and items, ensuring each has at least five associated interactions. 

\noindent \textbf{Metrics.} As shown in~\cite{zhong2024mirror,tang2018ranking}, higher-ranked positions in recommendation lists hold greater significance. So we utilize four commonly used metrics on Top-5 recommendations, including recall (REC), precision (PREC), mean average precision (MAP), and normalized discounted cumulative gain (NDCG), which emphasize different key aspects and complement each other in capturing user preferences and improving recommendation quality. Following previous works ~\cite{hu2025modality,jiang2024diffmm}, we split each user's interaction history into training, validation, and testing sets using an 8:1:1 ratio.

\noindent \textbf{Baselines.} We evaluate MRdIB on 6 representative MRSs: VBPR~\cite{he2016vbpr}, which incorporates visual signals through scalable matrix factorization; MMGCN~\cite{wei2019mmgcn}, which constructs modality-specific bipartite graphs for learning fine-grained user preferences; DualGNN~\cite{wang2021dualgnn}, which utilizes dual graph structures to discover user-specific fusion patterns; MGCN~\cite{yu2023multi}, which leverages modal-specific graphs for preference learning and representation fusion; FREEDOM~\cite{zhou2023tale}, which focuses on denoising user-item interactions while maintaining item relationships; and SOIL~\cite{su2024soil}, which enhances recommendation through second-order interest mining.

\noindent \textbf{Implementation Details.} In our study, we apply both the baseline methods and our proposed approach within the MMRec framework~\cite{zhou2023mmrec}. To ensure fair comparisons, we retain the standard setting for all baselines~\cite{yu2025mind} and utilize the pre-extracted visual and textual features provided by MMRec. We use Recall@5 on the validation set as the criterion for early stopping. We set the embedding size for both users and items to 64 across all models. All embedding parameters are initialized using the Xavier initialization~\cite{glorot2010understanding}, and model optimization is performed using the Adam optimizer~\cite{kinga2015method}. 
In line with the configurations adopted by recent studies on MRSs~\cite{yang2024multimodal, qi2025seeing}, we perform a grid search on hyperparameters $\alpha_1$, $\alpha_2$, and $\alpha_3$, with a detailed analysis provided in the 'Influence of Hyperparameters. 

\subsection{Overview Performance} 
In~\Cref{tab:performance_comparison}, we conduct an extensive evaluation of MRdIB on three datasets based on multiple baseline models. The results lead to two key observations:
\textbf{(1) MRdIB consistently and significantly enhances all baseline models.} On average, models with MRdIB improved by 8.47\% in REC@5 and 8.27\% in NDCG@5. The impact is particularly notable on simpler models like VBPR (up to 27.23\% gain) and remains effective even on the SOTA model SOIL (average 4.87\% gain in REC@5). This demonstrates that MRdIB's principled approach—filtering noise while disentangling unique, redundant, and synergistic information—effectively guides models to learn more powerful representations.
\textbf{(2) The performance gains from MRdIB are robust across different data domains.} The framework delivered strong and consistent improvements on the Baby, Sports, and Clothing datasets. This highlights that MRdIB provides a fundamental, domain-agnostic solution to the prevalent challenges of information redundancy and noise, making it a versatile and generalizable enhancement for various MRSs.

\begin{figure*}[t]
    \centering
    \subfigure[MGCN-$\alpha_1$\label{fig:mmgcn_beta1}]{
       \includegraphics[width=0.23\linewidth]{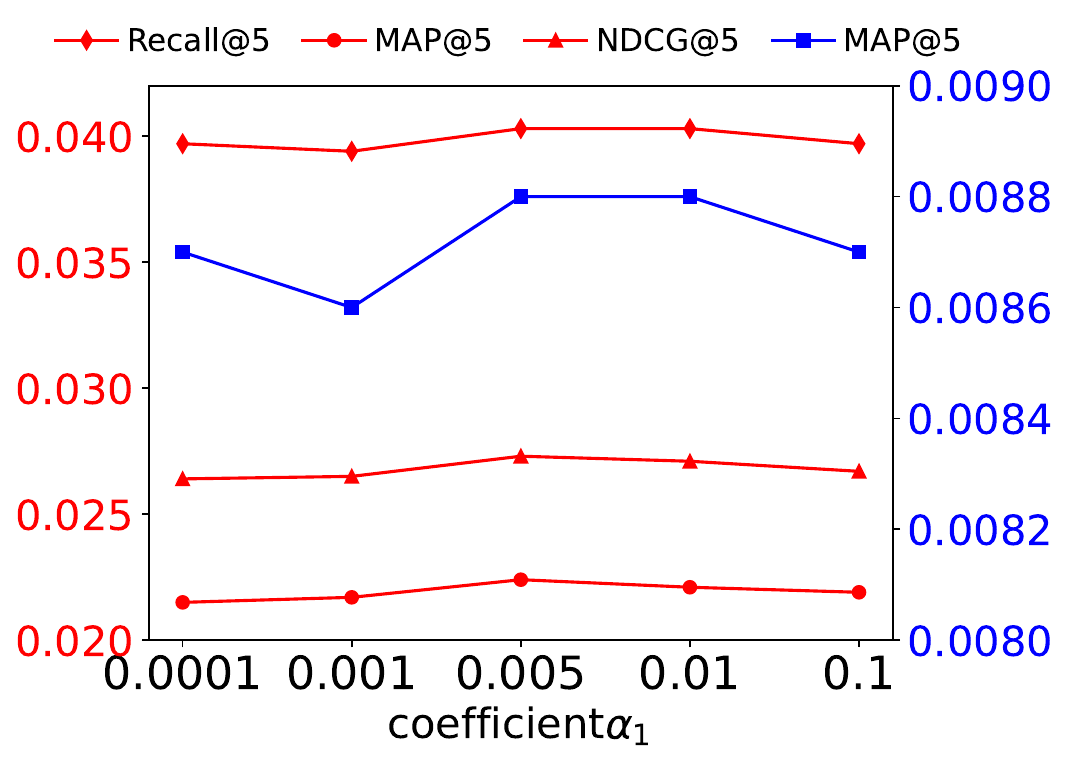}
      }
     \subfigure[MGCN-$\alpha_2$\label{fig:mmgcn_beta2}]{
      \includegraphics[width=0.23\linewidth]{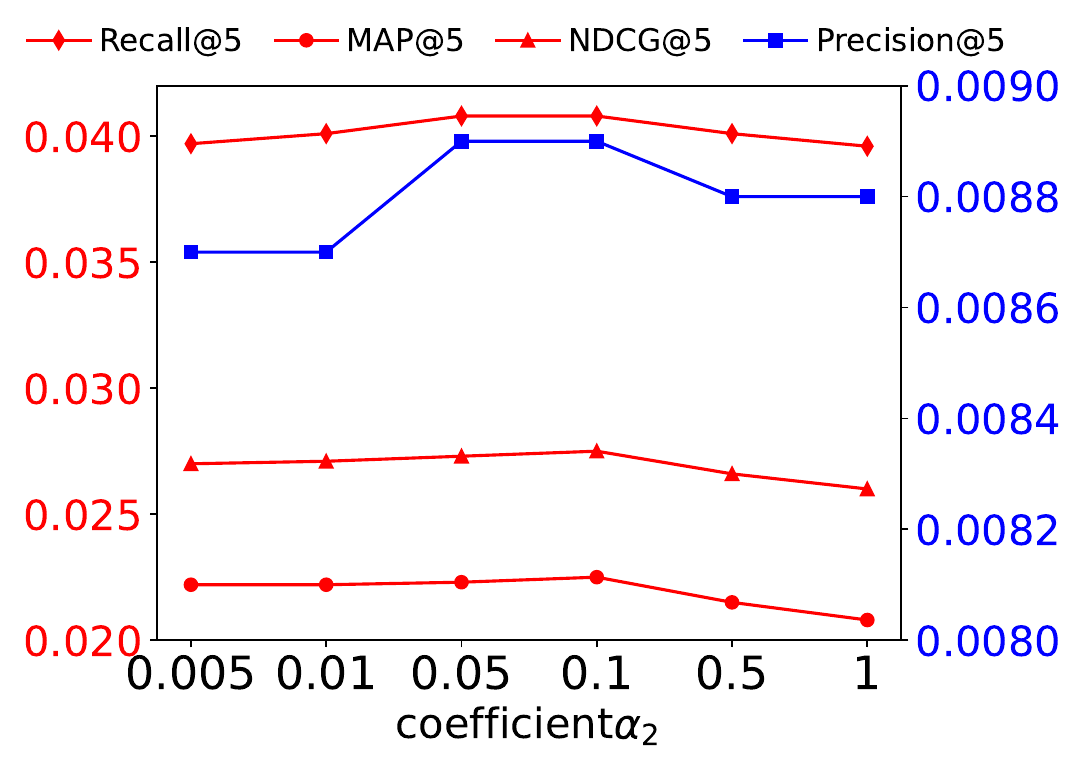}
     }
      \subfigure[MGCN-$\alpha_3$\label{fig:mmgcn_beta3}]{
       \includegraphics[width=0.23\linewidth]{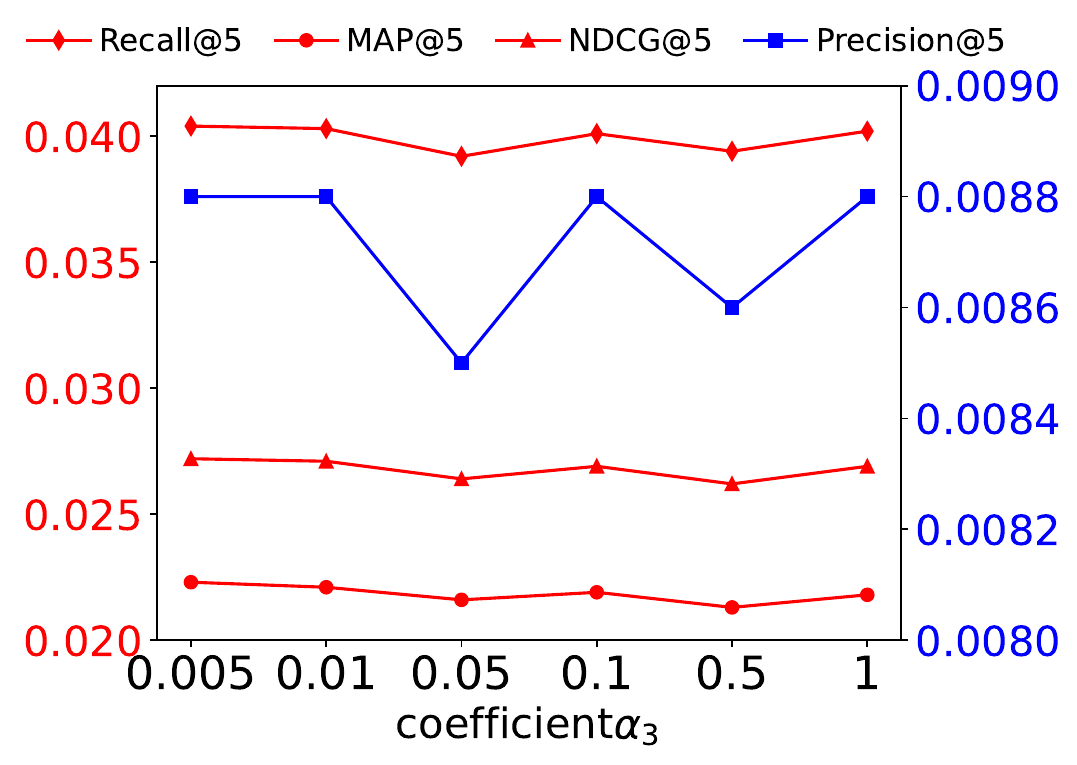}
    }
     \subfigure[Without MRdIB\label{fig:woCdMIB}]{
      \includegraphics[width=0.19\linewidth]{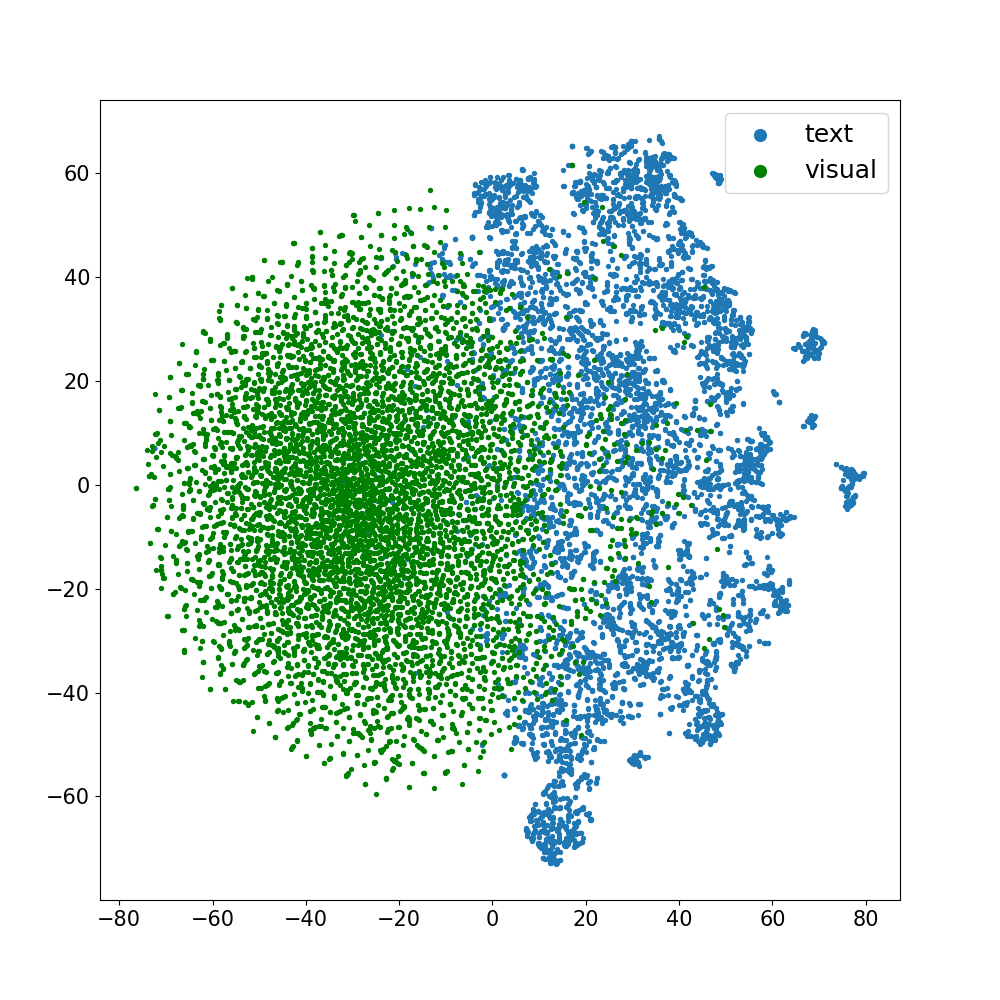}
    }
    \subfigure[FREEDOM-$\alpha_1$\label{fig:freedom_beta1}]{
      \includegraphics[width=0.23\linewidth]{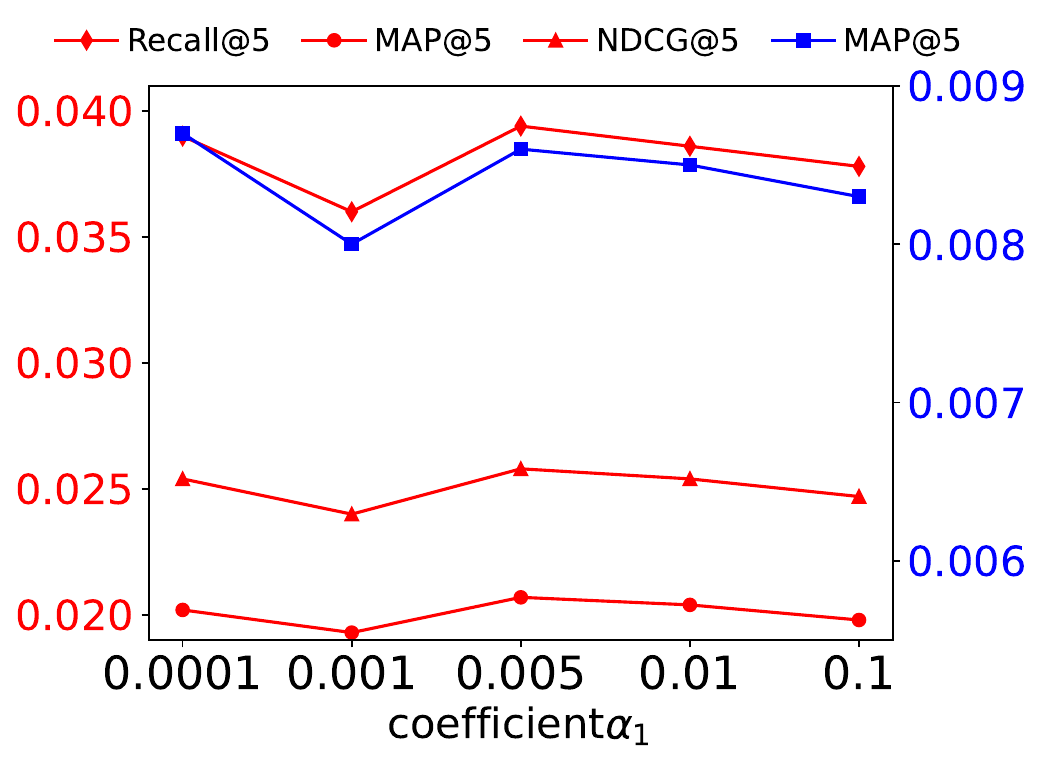}
    }
    \subfigure[FREEDOM-$\alpha_2$\label{fig:freedom_beta2}]{
      \includegraphics[width=0.23\linewidth]{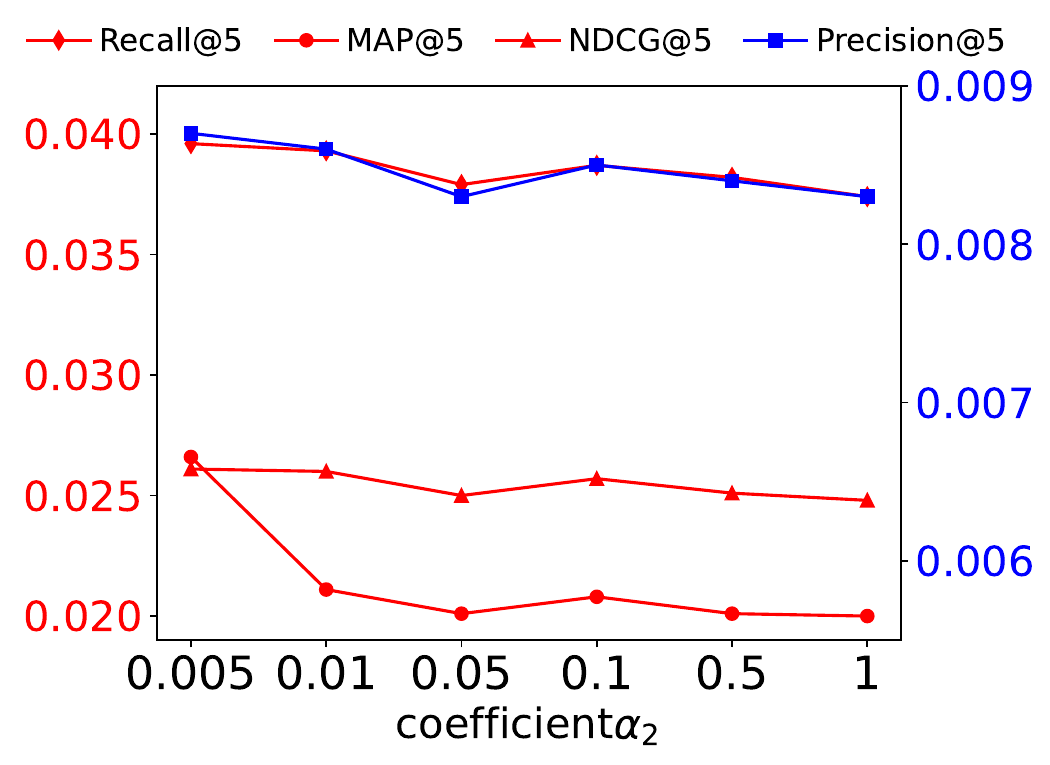}
    }
    \subfigure[FREEDOM-$\alpha_3$\label{fig:freedom_beta3}]{
      \includegraphics[width=0.23\linewidth]{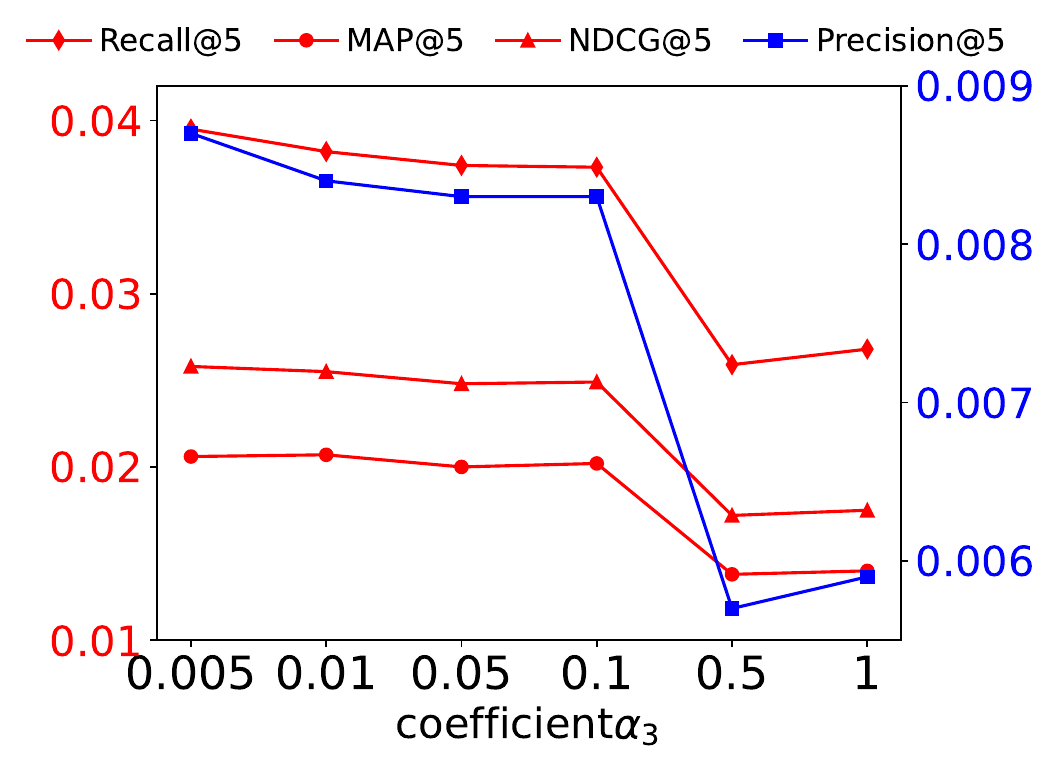}
    }
    \subfigure[With MRdIB\label{fig:wCdMIB}]{
      \includegraphics[width=0.19\linewidth]{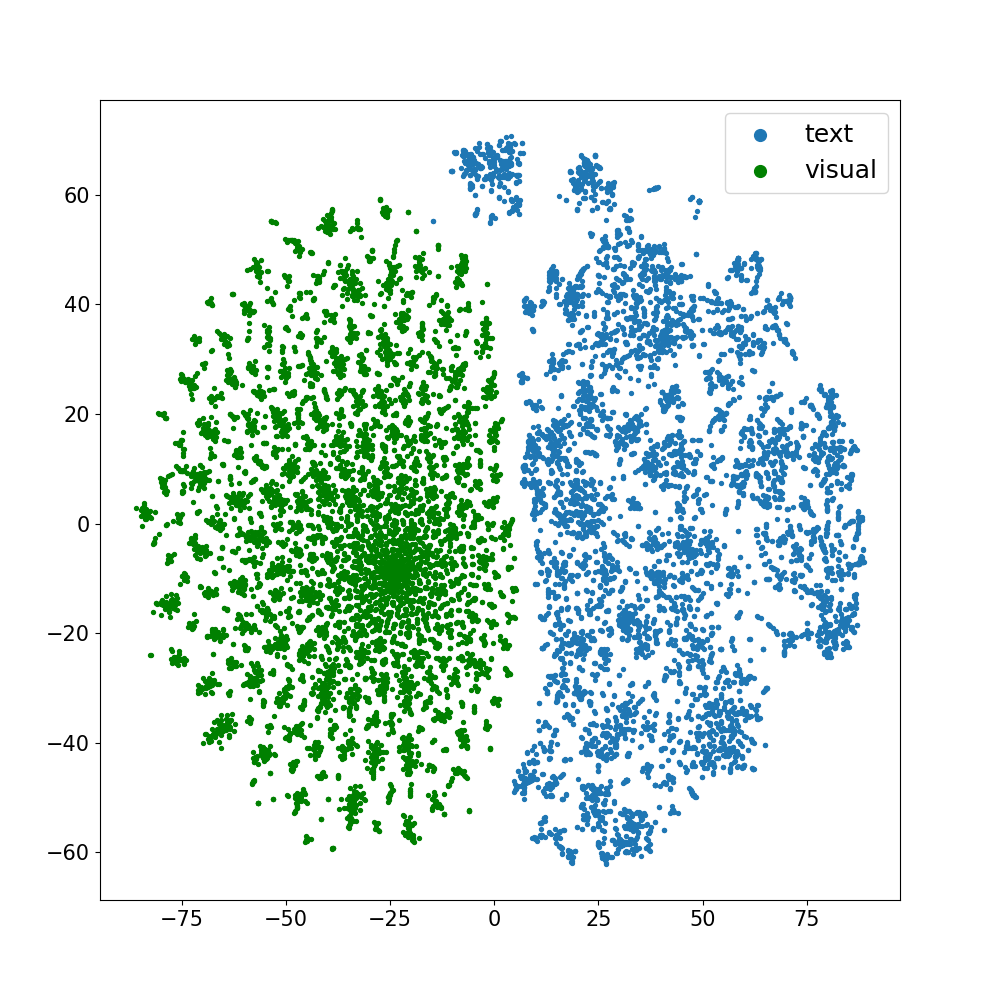}
    }
    \vspace{-3mm}
    \caption{(a-c, e-g) Sensitivity analysis of Top-5 recommendation accuracy for MGCN and FREEDOM with different coefficients on the Baby dataset. (d, h) The t-SNE visualization of modality representations learned in SOIL on the Sports dataset.}
    \vspace{-5mm}
    \label{fig:hyper}

\end{figure*}

\subsection{Ablation Study} 

To evaluate each component's contribution to MRdIB, we perform the following, with the results presented in \Cref{tab:ablation}:
\textbf{(1) Without MIB's compression constraint ($\alpha_1^-,\alpha_2,\alpha_3$ variant)}. Removing the compression constraint (the KL-divergence term in \Cref{eq:mib_loss}) consistently degrades performance across all metrics. The impact is particularly significant on SOTA models like SOIL and FREEDOM, which see performance drops of up to 6.31\% and 4.7\%, respectively. This underscores the importance of this constraint in filtering out irrelevant information, which is a critical step toward learning effective multimodal representations.
\textbf{(2) Without redundant information disentangling ($\alpha_1,$ $\alpha_2^-,$ $\alpha_3$ variant)}. Removing the objective that minimizes redundant information leads to a notable drop in performance. For example, MGCN's performance falls by an average of 5.13\% across all datasets. This underscores the importance of the redundant information disentanglement mechanism in MRdIB, which prevents the model from encoding overlapping information and allows for a more effective fusion of the disentangled representations.
\textbf{(3) Without unique information learning ($\alpha_1,\alpha_2,\alpha_3^-$ variant)}. Disabling the unique information learning objective ($\alpha_3=0$) leads to a clear drop in performance across all models. This result underscores the importance of preserving modality-specific signals. By ensuring that critical information unique to each modality is not lost during compression, this component enables the model to make more accurate and personalized recommendations.
\textbf{(4) Only MIB ($\alpha_1,\alpha_2^-,\alpha_3^-$ variant)}. When MRdIB is reduced to only the MIB framework, performance drops, as MIB alone cannot disentangle unique, redundant, and synergistic information. Nevertheless, models like SOIL, MGCN, and FREEDOM under this MIB-only setting still outperform their original versions, demonstrating the foundational benefit of filtering irrelevant information.
In summary, the ablation study confirms that all components of MRdIB are essential. Removing any single objective consistently degrades performance, validating our integrated design for learning disentangled representations.

\subsection{Influence of Hyperparameters } 
To guide hyperparameter selection and evaluate MRdIB's sensitivity, using the Baby dataset with MGCN and FREEDOM as baseline models, we conduct experiments exploring a range of values for three loss coefficients ($\alpha_1, \alpha_2, \alpha_3$): ${0.0001, 0.001, 0.005, 0.01, 0.05}$. Our experiments reveal varying levels of parameter sensitivity across different models, precise tuning of $\alpha_1$, $\alpha_2$, and $\alpha_3$ is essential for achieving an optimal balance. MGCN exhibits robust performance with minimal sensitivity to parameter variations, achieving optimal results at $\alpha_1=0.005$, $\alpha_2=0.05$, and $\alpha_3=0.005$. In contrast, FREEDOM shows greater performance fluctuations with hyperparameter changes, with optimal settings at $\alpha_1=0.0001$, $\alpha_2=0.005$, and $\alpha_3=0.005$. These distinct sensitivity patterns highlight the critical role of model-specific hyperparameter tuning in optimizing MRdIB's performance.

\subsection{Representation Disentanglement Analysis} 
To visually verify MRdIB's ability to disentangle representations, we visualize the unimodal representations ($Z_1$ and $Z_2$) learned by the SOIL model, both with and without our framework, using t-SNE on the Sports dataset. As shown in~\Cref{fig:woCdMIB}, the significant overlap between the baseline model's visual and textual representations reveals its failure to disentangle complex multimodal inputs, resulting in an inability to capture modality-unique information and resolve informational redundancy. 
In contrast,~\Cref{fig:wCdMIB} clearly shows that MRdIB effectively disentangles the representations, forcing them into distinct, well-separated clusters. This visualization provides strong evidence that our framework minimizes informational redundancy, as guided by the \Cref{eq:Rdobject}. This disentanglement allows the model to learn more discriminative representations that better leverage modality-unique information, ultimately improving recommendation performance.

\begin{table}[t]
\centering
\Huge
\resizebox{1\linewidth}{!}{
\begin{tabular}{lcccccc}
\hline
 & \textbf{VBPR} & \textbf{MMGCN} & \textbf{DualGNN} & \textbf{FREEDOM} & \textbf{MGCN} & \textbf{SOIL} \\
\hline
w/o MRdIB & 4.84 & 46.87 & 167.29 & 6.56 & 10.28 & 10.61 \\
w/ MRdIB    & 4.96 & 50.44 & 173.65 & 6.89 & 10.82 & 10.93 \\
\hline
\end{tabular}
}
\vspace{-3mm}
\caption{Training time (seconds) per epoch on Baby.}
\label{tab:efficiency}
\vspace{-5mm}
\end{table}

\subsection{More Analysis}
\subsubsection{Training Efficiency Analysis}

We analyzed the computational overhead of MRdIB on the Baby dataset. The primary overhead comes from the additional loss computations during training. As shown in~\Cref{tab:efficiency}, this results in a modest and controllable increase in training time, ranging from 3-8\%. Crucially, there is virtually no inference cost. All auxiliary components for disentanglement are discarded during prediction, and the final item representations can be pre-computed offline. Therefore, MRdIB's inference speed is comparable to the baseline models. We believe this small, one-time training cost is a valuable trade-off for the performance improvement (about 8\%).

\subsubsection{Comparison with other IB-based Methods}


Recent IB-based methods in recommendation~\cite{mai2022multimodal, yang2025less, zhao2025dvib} primarily focus on general denoising. While effective at filtering irrelevant information, they do not disentangle the structure of the remaining task-relevant information. The core novelty of MRdIB is its principled, fine-grained decomposition of this relevant information into unique, redundant, and synergistic components, guided by Partial Information Decomposition (PID). This allows for targeted optimization and distinguishes our work as a more precise, plug-and-play framework for representation learning. 
\section{Conclusion}


In this paper, we introduced MRdIB, a novel information-theoretic framework that addresses noise and information entanglement in MRSs. 
MRdIB first uses a Multimodal Information Bottleneck to filter irrelevant information, then applies a set of principled objectives guided by PID to disentangle the remaining signals into unique, redundant, and synergistic components. As a versatile plug-and-play module, experiments show MRdIB consistently and significantly enhances various MRS models. 
The primary limitation is the resource-intensive tuning of its three hyperparameters. Future work should focus on automating this balance to improve the framework's efficiency.

\bibliography{aaai2026}

\onecolumn 
\appendix
\section{Definitions}
The definition of mutual information of $X$ and $Y$:
\begin{equation}
    \begin{aligned}
        I(X; Y) &\equiv \mathbb{E} \left[ \log \frac{p(x, y)}{p(x)p(y)} \right] \\
        &= \int_{p(x,y)} p(x, y) \log \frac{p(x, y)}{p(x)p(y)} \, dxdy \\
        &= \mathbb{E}[\log p(x, y)] - \mathbb{E}[p(x)] - \mathbb{E}[p(y)] \\
        &= -H(X, Y) + H(X) + H(Y) \\
        &= H(Y) - H(Y|X) \\
        &= H(X) - H(X|Y)
    \end{aligned}
\end{equation}
where
\begin{equation}
    \begin{aligned}
    H(X)&=\mathbb{E}_{p(x)}[-\log p(x)]\\
    H(Y,X)&=\mathbb{E}_{p(x,y)}[-\log p(x,y)]\\
    &=-\int p(x,y)\log p(x,y)dxdy\\
        H(Y|X)&=\mathbb{E}_{p(x,y)}[-\log p(y|x)]
    \end{aligned}
\end{equation}
The definition of Kullback-Leibler(KL) divergence between two distributions is defined as:
\begin{equation}
    KL(p(x)||q(x))=\mathbb{E}_{p(x)}[\log \frac{p(x)}{q(x)}]
\end{equation}
\section{Proof}
\subsection{Derivation of MIB Loss}
\label{sec:proof1}
For multimodal learning, let the input modality information be represented by $x_1$ and $x_2$,and let the learned representations for these modalities be $z_1$ and $z_2$.The distribution of the modality representations satisfies the alignment condition of the multimodal information.
\begin{equation}
\begin{split}
    Z_1^*,Z_2^*=\arg\min_{Z_1,Z_2} I(X_1;Z_1) + I(X_2;Z_2) \\\text{s.t}. I(Z_1,Z_2;Y) = I(X_1,X_2;Y)
\end{split}
\end{equation}
The corresponding loss is as follows:
\begin{equation}
\begin{split}
    \mathcal{L}_{\text{MIB}} = \mathbb{E}_{p_\theta(x_{1},x_{2},y)}[-p_\theta(z_1,z_2|x_1,x_2)\log p_{\theta}(y\mid z_{1},z_{2})\\ + \alpha_1(KL(p_{\theta}(z_1|x_1)\log p_\theta (y|z_1)\parallel p(\mathcal{Z}))+KL(p_{\theta}(z_{2}\mid x_{2})\parallel p(\mathcal{Z})))]
\end{split}
\end{equation}
where $\alpha_1$ are positive constants.\\
{\bf Proof:} The multimodal learning information bottleneck can be summarized as:
\begin{equation}
    \begin{split}
    Z_1^*,Z_2^*=\arg\min_{Z_1,Z_2} I(X_1;Z_1) + I(X_2;Z_2) \\\text{s.t}. I(Z_1,Z_2;Y) = I(X_1,X_2;Y)
    \end{split}
\end{equation}
By applying the Lagrange multiplier method and noting that the mutual information between each input and the target variable remains constant (that is, $I(X_1; Y)$ and $I(X_2; Y)$ are fixed), the optimization problem can be reformulated as follows.
\begin{equation}
\begin{split}
Z_1^*,Z_2^*=\arg\min_{Z_1,Z_2}I(X_1;Z_1)+I(X_2;Z_2)-\beta I(Z_1,Z_2;Y)
\end{split}
\label{eq:mib}
\end{equation}
The $I(X_1;Z)$ are
\begin{equation}
    \begin{aligned}
I(X_1;Z_{1})& =\sum_{x_{1},z_{1}}p_{\theta}(x_{1},z_{1})\log\frac{p_{\theta}(z_{1}|x_{1})}{p_{\theta}(z_{1})} \\
&=\sum_{x_{1},z_{1}}p_{\theta}(x_{1},z_{1})\log p_{\theta}(z_{1}|x_{1})&-\sum_{x_{1},z_{1}}p_{\theta}(x_{1},z_{1})\log p_{\theta}(z_{1})
    \end{aligned}
\end{equation}
Replacing $p(z)$ with the prior distribution $P(\mathcal{Z})$, we obtain, according to Gibbs' inequality:
\begin{equation}
    \begin{split}
            &-\sum_{x_1,z_1} p_\theta\left(x_1, z_1\right) \log p_\theta\left(z_1\right)  \leq -\sum_{x_1,\mathcal{Z}} p_\theta\left(x_1, z_1\right) \log p(\mathcal{Z})
    \end{split}
\end{equation}
Further
\begin{equation}
    \begin{split}
I( X_1;Z_1) &\leq \sum_{x_1, z_1} p_\theta(x_1, z_1) \log p_\theta(z_1 | x_1) - \sum_{x_1, Z} p_\theta(x_1, z_1) \log p_\theta(Z) \\
&= \sum_{x_1, z_1, Z} p_\theta(x_1, z_1) \log \frac{p_\theta(z_1 | x_1)}{p_\theta(Z)} \\
&= \sum_{x_1, z_1, Z} p_\theta(x_1) p_\theta(z_1 | x_1) \log \frac{p_\theta(z_1 | x_1)}{p_\theta(Z)} \\
&= \mathbb{E}_{p_\theta(x_1)} \left[ KL\left( p_\theta(z_1 | x_1) \, \| \, p_\theta(Z) \right) \right]
\end{split}
\end{equation}
Based over equation, minizing $\mathbb{E}_{p_{\theta}(x_1)} \left[ KL\left( p_{\theta}(z_1 | x_1, x_2) \, \| \, p_{\theta}(Z) \right) \right]
$ is equel to minimize $I(Z_1;X_1)$, that means:
\begin{equation}
    \min \mathbb{E}_{p_{\theta}(x_1)} \left[ \beta KL\left( p_{\theta}(z_1 | x_1) \, \| \, p_{\theta}(Z) \right) \right] \propto \min I(Z_1; X_1)
\end{equation}
In the reasoning of the formula, we can similarly deduce that
\begin{equation}
    \begin{split}
            \min \mathbb{E}_{p_{\theta}(x_2)} \left[ \beta KL\left( p_{\theta}(z_2 | x_2) \, \| \, p_{\theta}(Z) \right) \right] \propto \min I(Z_2; X_2)
    \end{split}
\end{equation}
For $I(Z_1,Z_2;Y)$:
\begin{equation}
    \begin{split}
I\left(Z_1, Z_2; Y\right) & =\sum_{z_1,z_2, y} p_\theta\left(z_1, z_2,y\right) \log \frac{p_\theta\left(y \mid z_1,z_2\right)}{p_\theta(y)} \\
& =\sum_{z_1,z_2, y} p_\theta\left(z_1,z_2, y\right) \log p_\theta\left(y \mid z_1,z_2\right) -\underbrace{\sum_{z_1,z_2,y} p_\theta\left(z_1, z_2,y\right) \log p_\theta(y)}_{H(Y) \geq 0} \\
& \geq \sum_{z_1,z_2 y} p_\theta\left(z_1, z_2, y\right) \log p_\theta\left(y \mid z_1, z_2\right)
\end{split}
\end{equation}
Meanwhile,in the experiment,one $x_1$ that means one $z_1$,so we get:
\begin{equation}
    \begin{split}
        p_{\theta}(z_{1},z_2, y)=p_{\theta}(x_{1},x_2, y)\\
        p_{\theta}(y|z_{1},z_2)=p_{\theta}(y|x_{1},x_1)
    \end{split}
\end{equation}
Further:
\begin{equation}
    \begin{split}
I(Z_{1}, Z_2;Y) &\geq\sum_{y,x_{1},x_2}p_{\theta}(x_{1},x_2,y)\log p_{\theta}(y|x_{1},x_2) \\
&=\mathbb{E}_{p(y,x_1,x_2)}[\log p_\theta(y|x_1,x_2)] 
    \end{split}
\end{equation}
Maximizing $\mathbb{E}_{p(y,x_1,x_2)}[\log p_\theta(y|x_1,x_2)]$is equal to maximize $I(Z_{1},Z_2;Y)$
that means
\begin{equation}
    \begin{split}
        \max\mathbb{E}_{p(y,x_1,x_2)}[\log p_\theta(y|z_1,z_1)]\propto\max\beta I(Z_1,Z_2;Y)
    \end{split}
\end{equation}
 Based on the above steps, we obtain:
 \begin{equation}
     \begin{split}
Z_1^*,Z_2^*&=\arg\arg\min_{Z_1,Z_2}I(X_1;Z_1)+I(X_2;Z_2)-\alpha_1 I(Z_1,Z_2;Y)\\
&=\arg\min_{\theta}\mathbb{E}_{p_\theta(x_{1},x_{2},y)}[-p_\theta(z_1,z_2|x_1,x_2)\log p_{\theta}(y\mid z_{1},z_{2})\\ &+ \alpha_1 (KL(p_{\theta}(z_{1}\mid x_{1})\parallel p(\mathcal{Z}))+KL(p_{\theta}(z_{2}\mid x_{2})\parallel p(\mathcal{Z})))]
     \end{split}
 \end{equation}
Here, $\alpha_1$ are constants, which result from redefinition or normalization to fit the optimization framework. And then we have:
\begin{equation}
\begin{split}
    \mathcal{L}_{\text{MIB}} = \mathbb{E}_{p_\theta(x_{1},x_{2},y)}[-p_\theta(z_1,z_2|x_1,x_2)\log p_{\theta}(y\mid z_{1},z_{2}) + \alpha_1 (KL(p_{\theta}(z_{1}\mid x_{1})\parallel p(\mathcal{Z}))+KL(p_{\theta}(z_{2}\mid x_{2})\parallel p(\mathcal{Z})))]
\end{split}
\end{equation}
\subsection{Modality Representation Alignment} 
Different modalities have distinct distributions, and directly merging them can lead to information loss and adversely affect performance. Therefore, it is essential to align the distributions of modality representations first. Previous work has designed regularizers to reduce the distance between different representations. We minimize the differences in modality representation distributions by aligning multiple modalities to the same prior distribution. The specific constraints are given by the following expression:
\begin{equation}
    \begin{split}
        \theta^{*}=\arg\min_{\theta}\mathbb{E}_{p_\theta(x_{1},x_{2},y)}[-p_\theta(z_1,z_2|x_1,x_2)\log p_{\theta}(y\mid z_{1},z_{2})]\\s.t. \left\{\begin{matrix}KL(p_\theta(z_1\mid x_1)\parallel p(\mathcal{Z}))=0\\KL(p_\theta(z_2\mid x_2)\parallel p(\mathcal{Z}))=0\end{matrix}\right.
    \end{split}
\end{equation}
Applying the Lagrange method, we equivalently optimize the following objective:
\begin{equation}
    \begin{split}
    \theta^{*}=\arg\min_{\theta}\mathbb{E}_{p_\theta(x_{1},x_{2},y)}[-p_\theta(z_1,z_2|x_1,x_2)\log p_{\theta}(y\mid z_{1},z_{2})\\ + \alpha_1(KL(p_{\theta}(z_{1}\mid x_{1})\parallel p(\mathcal{Z}))+KL(p_{\theta}(z_{2}\mid x_{2})\parallel p(\mathcal{Z})))]
    \end{split}
\end{equation}
According to the derivation in ~\Cref{sec:proof1}, we have the following theorem:
\textbf{Theorem 1} For multimodal learning, let the input modality information be $x_1$ and $x_2$, and let the corresponding learned modality representations be $z_1$ and $z_2$. The alignment of modality representation distributions is equivalent to optimizing the multimodal information bottleneck.
\begin{equation}
\begin{split}
Z_1^*,Z_2^*=&\arg\min_{Z_1,Z_2}I(X_1;Z_1)+I(X_2;Z_2)-\beta I(Z_1,Z_2;Y)
\\=&\arg\min_{\theta}\mathbb{E}_{p_\theta(x_{1},x_{2},y)}[-p_\theta(z_1,z_2|x_1,x_2)\log p_{\theta}(y\mid z_{1},z_{2})\\ +& \alpha_1(KL(p_{\theta}(z_{1}\mid x_{1})\parallel p(\mathcal{Z}))+KL(p_{\theta}(z_{2}\mid x_{2})\parallel p(\mathcal{Z})))]
\end{split}
\end{equation}
\subsection{Derivation of MRdIB loss}
\label{sec:proof2}
By combining MIB with Unique Causal Information Learning, Redundant Causal Information Disentangling, and Synergistic Causal Information Learning, the following constraints for MCdIB can be obtained:
\begin{equation}
    \begin{split}
        Z_1^*,Z_2^* \arg&\min_{Z_1,Z_2}I(X_1;Z_1)+I(X_2;Z_2)\\s.t.&\min I(Z_1;Z_2)\\& \max I(Z_1,Z_2;Y),\\&\max (I(Y;Z_1,Z_2))
    \end{split}
\end{equation}
This can be further optimized as:
\begin{equation}
    \begin{split}
        Z_1^*,Z_2^*= \arg\min_{Z_1,Z_2}I(X_1;Z_1)+I(X_2;Z_2)+\beta_1I(Z_1;Z_2) -\beta_2I(Z_1,Z_2;Y)-\beta_3(I(Y;Z_1,Z_2))
    \end{split}
\end{equation}
The optimization objective of CdMIB is:
\begin{equation}
\begin{split}
    Z_1^*,Z_2^*=&\arg\min_{Z_1,Z_2}E_{p(x_1,x_2,y)}[\alpha_1(KL(p_\theta(z_1\mid x_1) \\&+KL(p_\theta(z_2\mid x_2)))+\alpha_2(E_{p_\theta(z_1)p_{\theta}(z_2)}[f]\\&-\log E_{p_\theta(z_1)p_\theta(z_2)}[e^f])-\log p_\theta(y\mid z_1,z_2)\\&-\alpha_3(p_\theta(z_1\mid x_1)\log p_\theta(y\mid z_1)\\&+p_\theta(z_2\mid x_2)\log p_\theta (y\mid z_2))]
\end{split}
\label{eq:object}
\end{equation}
So we get:
\begin{equation}
\begin{aligned}
    & \mathcal{L}_{\text{MCdIB}} = \mathbb{E}_{p(x_1,x_2,y)}[ -p_\theta(z_1,z_2|x_1,x_2)\log p_\theta(y\mid z_1,z_2) \\
    & \qquad +\alpha_1(KL(p_\theta(z_1\mid x_1)\parallel p(\mathcal{Z}))+KL(p_\theta(z_2\mid x_2)\parallel p(\mathcal{Z})) \\
    & \qquad +\alpha_2(E_{p_\theta(z_1,z_2)}[f]-\log E_{p_\theta(z_1)p_\theta(z_2)}[e^{f}]) \\
    & \qquad +\alpha_3(-p_\theta(z_1\mid x_1)\log p_\theta(y\mid z_1)-p_\theta(z_2\mid x_2)\log p_\theta (y\mid z_2))]
\end{aligned}
\label{eq:object}
\end{equation}
\textbf{Proof}: For $I(Z_1; Z_2)$, according to the definition of mutual information and KL divergence, we have
\begin{equation}
    \begin{split}
        I(Z_1;Z_2) &= \sum_{z_2}\sum_{z_1} p(z_2|z_1)p(z_1)\log\frac{p(z_2|z_1)}{p(z_2)}
        \\&=\sum_{z_2}\sum_{z_1} p(z_2|z_1)p(z_1)\log\frac{p(z_2|z_1)p(z_1)}{p(z_2)P(z_1)}
        \\&=\sum_{z_2}\sum_{z_1} p(z_2,z_1)\log\frac{p(z_2,z_1)}{p(z_2)P(z_1)}
        \\& = KL(p(z_1,z_2) \parallel p(z_1)p(z_2))
    \end{split}
\end{equation}
Let $\tilde{p}(z_1, z_2) := \frac{1}{\phi} e^f p_\theta(z_1) p_\theta(z_2)$,
where $\phi = \mathbb{E}_{p_\theta(z_1) p_\theta(z_2)}[e^f]$ and $f$ is a discriminant network of multi-layer perceptrons. Obviously, we have:
\begin{equation}
    \begin{split}
\mathbb{E}_{p_\theta(z_1, z_2)}[f] - \log \phi &= \mathbb{E}_{p_\theta(z_1, z_2)}\left[\log e^f - \log \phi\right]\\&
= \mathbb{E}_{p_\theta(z_1, z_2)}\left[\log \frac{e^f}{\phi}\right]\\&
= \mathbb{E}_{p_\theta(z_1, z_2)}\left[\log \frac{\tilde{p}(z_1, z_2)}{p_\theta(z_1)p_\theta(z_2)}\right]
    \end{split}
\end{equation}
Further:
\begin{equation}
    \begin{split}
        KL(p_\theta(z_1, z_2) \parallel p_\theta(z_1)p_\theta(z_2)) - \left(\mathbb{E}_{p_\theta(z_1, z_2)}[f] - \log \phi\right)
        =& \mathbb{E}_{p_\theta(z_1, z_2)}\left[\log \frac{p_\theta(z_1, z_2)}{p_\theta(z_1)p_\theta(z_2)}\right] - \mathbb{E}_{p_\theta(z_1, z_2)}[f] \\
        =& \mathbb{E}_{p_\theta(z_1, z_2)}\left[\log \frac{p_\theta(z_1, z_2)}{\tilde{p}(z_1, z_2)}\right] \\
        =& KL(p_\theta(z_1, z_2) \parallel \tilde{p}(z_1, z_2))
    \end{split}
\end{equation}
Because KL divergence is non-negative, we have:
\begin{equation}
    \begin{split}
        &KL(p_\theta(z_1, z_2) \parallel p_\theta(z_1)p_\theta(z_2)) - \left(\mathbb{E}_{p_\theta(z_1, z_2)}[f] - \log \phi\right) \geq 0 \\
        &\Rightarrow KL(p_\theta(z_1, z_2) \parallel p_\theta(z_1)p_\theta(z_2)) \geq \left(\mathbb{E}_{p_\theta(z_1, z_2)}[f] - \log \phi\right) \\
        &\quad\geq \left(\mathbb{E}_{p_\theta(z_1, z_2)}[f] - \log \mathbb{E}_{p_\theta(z_1)p_\theta(z_2)}[e^f]\right)
    \end{split}
\end{equation}
Therefore, we can obtain:
\begin{equation}
    \begin{split}
\min I(Z_1; Z_2) \quad \propto \min &\mathbb{E}_{p_\theta(z_1, z_2)}[f]  - \log \mathbb{E}_{p_\theta(z_1)p_\theta(z_2)}[e^f]
    \end{split}
\end{equation}
So we have:
\begin{equation}
    \begin{split}
    \mathcal{L}_{\Delta I^U_{Z_1 \to Y}, \Delta I^U_{Z_2 \to Y}} = &\mathbb{E}_{p_\theta(x_1, y)} [-p_\theta(z_1|x_1)\log p_\theta\left(y \mid z_1\right)] \\+& \mathbb{E}_{p_\theta(x_2, y)} [-p_\theta(z_2|x_2)\log p_\theta\left(y \mid z_2\right)]\\
    \end{split}
    \label{eq:unique_loss}
\end{equation}
According to the process of \Cref{sec:proof1}, we have:
\begin{equation}
    \begin{split}
\min \left\{\begin{array} { l } 
{ I ( X _ { 1 }; Z _ { 1 } ) } \\
{ I ( X _ { 2 } ; Z _ { 2 } ) }
\end{array} \quad \propto \operatorname { m i n } \left\{\begin{array}{l}
K L\left(p_\theta\left(z_1 \mid x_1, x_2\right) \| p(\mathcal{Z})\right) \\
K L\left(p_\theta\left(z_2 \mid x_1, x_2\right) \| p(\mathcal{Z})\right)
\end{array}\right.\right.
    \end{split}
\end{equation}

\begin{equation}
    \begin{split}
\max \left\{\begin{array} { l } 
{ I ( Y ; Z _ { 1 } ) } \\
{ I ( Y ; Z _ { 2 } ) }
\end{array} \quad \propto \operatorname { m i n } \left\{\begin{array}{l}
-p_\theta(z_1\mid x_1)\log p_\theta(y\mid z_1)\\
-p_\theta(z_2\mid x_2)\log p_\theta (y\mid z_2))
\end{array}\right.\right.
    \end{split}
\end{equation}
\begin{equation}
    \begin{split}
        \max I(Z_1,Z_2;Y) \quad \propto \max\mathbb{E}_{z_1,x_2,y}[\log p_\theta(y|z_1,z_2)]
    \end{split}
\end{equation}
We have:
\begin{equation}
    \begin{split}
    \mathcal{L}_{\Delta I^U_{Z_1 \to Y}, \Delta I^U_{Z_2 \to Y}} = &\mathbb{E}_{p_\theta(x_1, y)} [-p_\theta(z_1|x_1)\log p_\theta\left(y \mid z_1\right)] + \mathbb{E}_{p_\theta(x_2, y)} [-p_\theta(z_2|x_2)\log p_\theta\left(y \mid z_2\right)]\\
    \end{split}
    \label{eq:unique_loss}
\end{equation}
\begin{equation}
\label{eq:disentangle}
    \begin{split}
        \mathcal{L}_{\Delta I^S_{Z_1, Z_2 \to Y}} = \mathbb{E}_{p(x_1,x_2,y)}[-p_\theta(z_1,z_2|x_1,x_2)\log p_\theta(y|z_1,z_2)]
    \end{split}
\end{equation}
Based on the above derivation, we have:
\begin{equation}
\begin{aligned}
    & \mathcal{L}_{\text{MCdIB}} = \mathbb{E}_{p_\theta(x_1,x_2,y)}[ -p_\theta(z_1,z_2|x_1,x_2)\log p_\theta(y\mid z_1,z_2) \\
    & \qquad +\alpha_1(KL(p_\theta(z_1\mid x_1)\parallel p(\mathcal{Z}))+KL(p_\theta(z_2\mid x_2)\parallel p(\mathcal{Z})) \\
    & \qquad +\alpha_2(E_{p_\theta(z_1,z_2)}[f]-\log E_{p_\theta(z_1)p_\theta(z_2)}[e^{f}]) \\
    & \qquad +\alpha_3(-p_\theta(z_1\mid x_1)\log p_\theta(y\mid z_1)-p_\theta(z_2\mid x_2)\log p_\theta (y\mid z_2))]
\end{aligned}
\label{eq:object}
\end{equation}


\end{document}